\documentclass{article}  
\usepackage{jamaica04}
\usepackage{graphicx}
\frompage{000} \topage{000}                                              %|
\def\rightmark{BRAHMS results}
\def\leftmark{R. Debbe et al.}
 
\title{BRAHMS results in the context of saturation and quantum evolution  
} 
\authors{
{R. Debbe$^1$ , for the BRAHMS Collaboration %
}\\[2.812mm]
{\normalsize
\hspace*{-8pt}$^1$ Brookhaven National Laboratory, \\ 
Upton, NY 11973 \\[0.2ex] 
}
}
 
\abstract{We report BRAHMS results from RHIC d+Au and p+p collisions at $\sqrt{s_{NN}}=200GeV$. A remarkable change in
the nuclear modification factor $R_{dAu}$ is seen as the pseudorapidity of the detected charged hadrons changes from
zero at mid-rapidity to 3.2 at the most forward angle studied during the 2003 run. For pseudorapidity $\eta>1$ the suppression of the
$R_{cp}$ factor is more pronounced in the sample of central events in contrast to the behavior at mid-rapidity
where the central events show higher enhancement compared to a semi-central sample. These results are consistent
with a saturated Au wave function strongly affected by quantum evolution at higher values of rapidity.}

\keyword{RHIC, d+Au, p+p, saturation, quantum evolution} 
\PACS{25.75.Dw, 13.18.Hd, 25.75.-q}
 
\begin{document}
 
\maketitle
\setcounter{page}{1}

\section{Introduction}\label{intro}
The first BRAHMS results from d+Au collisions at forward rapidities have generated heated discussions since 
they were presented in their preliminary form at the DNP meeting in Tucson AZ. At that time, the theoretical 
work offered two clearly differentiated views. On one side, the groups that postulate the formation of the Color Glass Condensate CGC  
\cite{McLerranVenu} at RHIC, had results that demonstrated the presence of Cronin enhancements 
\cite{CroninEXP} in CGC \cite{JamalFirst,Jamal,Dumitru,Kovchegov} 
as well as studies that included quantum evolution to describe how the nuclear modification factor would be modified as the collisions are studied at 
higher rapidities \cite{KKT,KLM}  and \cite{Wiedemann}. These groups describe the d+Au collisions as coherent multiple interactions between the deuteron valence quarks and a saturated Au wave function at small values of {\it x}. As the density of gluons in the Au nuclei grows, higher order corrections to the 
gluon density are included within 
the formalism referred as quantum evolution. The net effect of these corrections is an overall reduction of the 
number of gluons compared to scaled p+p collisions. The effect of quantum evolution is also present in the 
centrality dependence; the more central the collisions, the stronger the effect of quantum evolution making
the suppression of the number of gluons more pronounced in central events.      

Other groups had worked the problem based on an standard description of the Cronin enhancement as incoherent 
multiple scattering
at the partonic level \cite{Vitev,XWang,Accardi}. This description of the d+Au collisions would continue exhibiting a Cronin type enhancement close to the deuteron fragmentation region, and the strength of the enhancement
would increase with the centrality of the collision.  

We present here a description of the BRAHMS results \cite{RdApaper} within the context of saturation that
includes the effects of quantum evolution. We are well aware that even though our results are consistent with
the descriptions offered by the theory that includes the presence of saturation in the initial state, more
measurements are necessary to eliminate other explanations.

\section{Experimental results}\label{Experimental}  

The spectra presented in this contribution were extracted from data collected with both BRAHMS spectrometers, the mid-rapidity 
spectrometer (MRS) and the 
front section of the forward spectrometer (FFS). 
A detailed description of the BRAHMS experimental setup can be found in \cite{BRAHMSNIM}. 
The low multiplicity
of charged particles in the proton+proton and d+Au collisions required an extension of the basic apparatus with a set of
scintillator counters (called INEL detectors).  These detectors \cite{BRAHMS_daumult} cover pseudo-rapidities in the
range: $3.1 \leq \mid \eta \mid \leq 5.29$, and  
define a minimum biased trigger.
This trigger is estimated to select $\approx 91\% \pm 3\%$ of the 2.4 barns d+Au 
inelastic cross section and $71\% \pm 5\%$ of the total inelastic proton-proton 
cross section of 41 mb. 
The INEL detector  
was also used to select events with collision 
vertex within $\pm 15$ cm of the nominal collision point with a resolution of 5 cm.

The centrality of the collision was extracted from the multiplicity of the event measured within the
angular region $\mid \eta \mid \leq 2.2$ with a combination of silicon and scintillator counters \cite{BRAHMS_Tiles}.

\subsection{Spectra} \label{Spectra}

\begin{figure}[htb]
\vspace*{-.5cm}
                 \insertplot{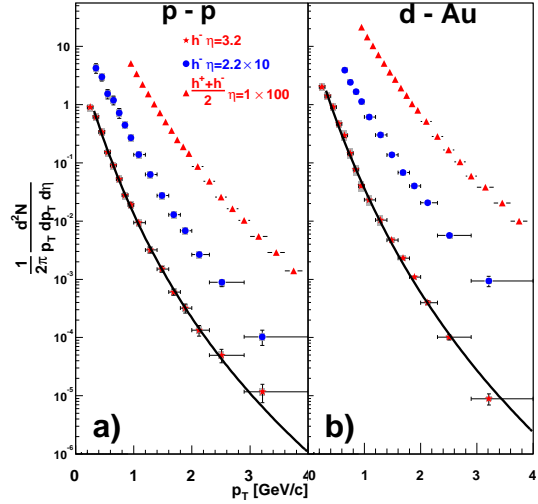}
%\vspace*{-2cm}
\caption[]{Spectra for charged hadrons at different 
pseudo-rapidities. Panel a shows the spectra obtained from proton-proton collisions and panel b those from d+Au collisions. 
The top  most distributions in both panels correspond to the invariant yields of $\frac{h^+ + h^-}{2}$ measured at
40 degrees with the MRS spectrometer (scaled by 100 for clarity purposes), followed by the yields of negative hadrons measured at 12 (scaled up by 10) and 4 degrees respectively. More details about these distributions can
be found in Ref. \cite{RdApaper} }
\label{fig:spectra}
\end{figure}

Figure \ref{fig:spectra} shows the invariant yields obtained from p+p collisions (panel a) and d+Au collisions (panel b).
For each system we studied particle production at 40 degrees with the MRS spectrometer and 12 and 4 degrees with the 
FFS spectrometer. Each distribution was obtained from several magnetic field settings and corrected for the 
spectrometer acceptance, tracking and trigger efficiency. No corrections were applied to the spectra for absorption or weak decays. 
Statistical errors are shown as vertical lines, and an overall systematic error of 15\%  is assigned to  each  point. The p+p spectra have also been corrected for trigger efficiency by $13 \pm 5\% $ to make 
them minimum biased with respect to the total inelastic cross section.  
We fitted the spectra at $\eta = 3.2$ with a power law function 
$  \frac{C}{(1+\frac{p_{T}}{p_0})^n}$ and the integral of that 
function over $p_{T}^2$ is 
compared for consistency in table \ref{tab1} with UA5 results 
\cite{UA5} for the p+p system, as well as
our own multiplicity measurement for the d+Au system \cite{BRAHMS_daumult}.

\begin{table}[hb] 
\vspace*{-12pt}
\caption[]{Fits to power law shapes at $\eta=3.2$.}\label{tab1}
\vspace*{-14pt}
\begin{center}
\begin{tabular}{lcccc}
\hline\\[-10pt]
System & $\frac{dN}{d\eta}_{fit} / \frac{dN}{d\eta}_{meas}$ & $p_{0}$ & n &
$\chi^2/NDF$\\ 
& & GeV/c &  & \\
\hline\\[-10pt]
p + p & 1.05 $\pm$ 0.06/0.95 $\pm$ 0.07 & 1.18 $\pm$ 0.16  & 10.9 $\pm$ 0.9 & 
13. / 11 \\
d + Au & 2.23 $\pm$ 0.09 / 2.1 $\pm$ 0.6 & 1.52 $\pm$ 0.1 & 12.3 $\pm$ 0.5 & 102. / 11 \\ 

\hline 
\end{tabular}
\end{center}
\end{table}

\subsection{Nuclear modification factor $R_{dAu}$}\label{RdA}

The d+Au system is compared to a ``simpler'' one: p + p where in this particular case, we do not expect the
effects of saturation. This comparison is based on the assumption that 
the production of moderately high transverse momentum particles scales with the
number of binary collisions $N_{coll}$ in the initial stages. The so-called  nuclear modification 
factor is defined as:
\begin{equation}
R_{dAu} \equiv \frac{1}{N_{coll}} \frac{N_{dAu}(p_{T},\eta)}{N_{pp}(p_{T},\eta)}
\label{equation1}
\end{equation}
where $N_{coll} $ is estimated to be equal to $7.2 \pm 0.3$ for minimum bias collisions.

\begin{figure}[htb]
\vspace*{-.5cm}
                 \insertplot{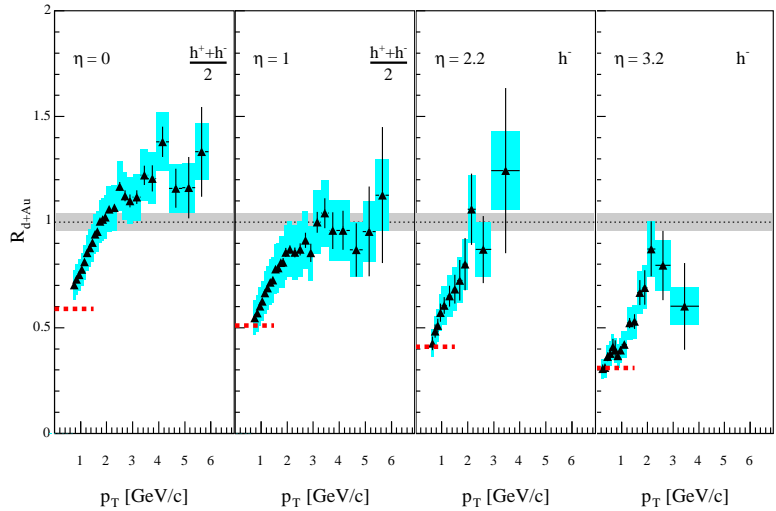}

\caption{\label{fig:ratio} Nuclear modification factor for charged
  hadrons at pseudorapidities $\eta=0,1.0,2.2,3.2$. Statistical
  errors are shown with error bars. Systematic
  errors are shown with shaded boxes 
  with widths set by the bin sizes.                                          
  The 
  shaded band around
  unity indicates the estimated error on the normalization to $\langle N_{coll} \rangle$. 
  Dashed lines at $p_T<1$ GeV/c show the normalized charged particle 
  density ratio $\frac{1}{\langle
  N_{coll}\rangle}\frac{dN/d\eta(d+Au)}{dN/d\eta(pp)}$.}

\end{figure}

Figure \ref{fig:ratio} shows the nuclear modification factor defined above for four $\eta$ values. 
At mid-rapidity ($\eta = 0$), the nuclear modification factor exceeds 1 for
transverse momenta greater than 2 GeV/c in a similar way as the measurements
performed by Cronin  at lower energies \cite{CroninEXP}. 
 
A shift of one unit of rapidity is enough to make the Cronin type enhancement disappear, and further 
increases in 
$\eta$ decrease even further the nuclear modification factor $R_{dAu}$.
Further details from this Figure \ref{fig:ratio} can be found in Ref. \cite{RdApaper}.

We see a one to one correspondence between the $R_{dAu}$ values at low $p_{T}$  and the ratio 
 $\frac{1}{\langle
N_{coll}\rangle}\frac{dN/d\eta(d+Au)}{dN/d\eta(pp)}$ as demonstrated in Fig. \ref{fig:ratio} where that ratio
is shown as dashed lines at $p_T <1$.  

\subsection{Centrality dependence $R_{cp}$}\label{Rcp}

In the context of saturation, the suppression of the overall number of gluons in the Au wave function depends
on a power of the number of participating nucleons $N^{Au}_{part}$; the higher this number, the stronger the 
suppression. We extract the number of participants from the multiplicity of the collision measured in the range
$\mid \eta \mid \leq 2.2$. To study the dependence on centrality or number of participants three data samples 
of different centralities 
were defined according to
the multiplicity of each event, and scaled histograms in transverse momentum were filled:
 $N_{central}(p_{T}) \equiv \frac{1}{N_{coll}}N_{0-20\%}(p_{T}) $ for  events with multiplicities ranging from 0
 to 20\%.
 $N_{semi-central}(p_{T}) \equiv \frac{1}{N_{coll}}N_{30-50\%}(p_{T}) $ for semi-central 
events with multiplicities ranging from 30 to 50\%, and finally, 
 $N_{periph}(p_{T}) \equiv \frac{1}{N_{coll}}N_{60-80\%}(p_{T}) $ for peripheral events
with multiplicities ranging from 60 to 80\% with $N_{coll}$ values listed in table \ref{tab2}.

With these histograms, two ratios were constructed: 
$R^{central}_{CP} = \frac{N_{central}(p_{T})} {N_{periph}(p_{T})}$ and 
$R^{semi-central}_{CP} = \frac{N_{semi-central}(p_{T})} {N_{periph}(p_{T})}$.

\begin{figure}[htb]
\vspace*{-.5cm}
                 \insertplot{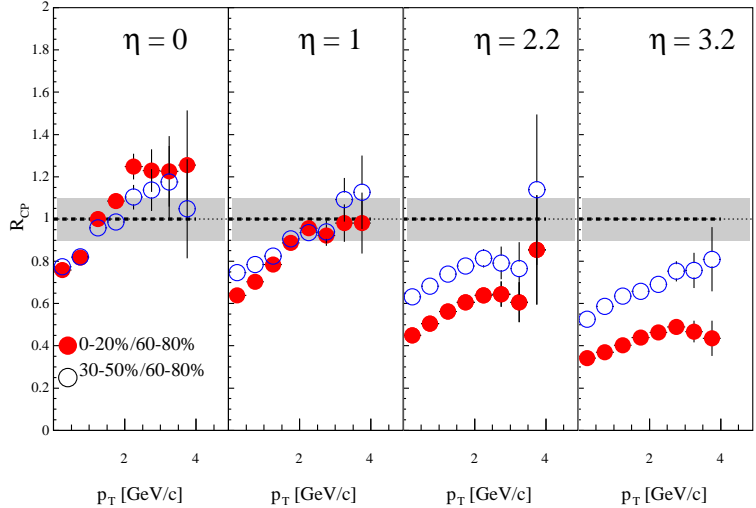}

\caption[]{ Central (full points) and
    semi-central (open points) $R_{cp}$ ratios (see text for details)
    at pseudorapidities $\eta=0,1.0,2.2,3.2$. Systematic errors ($\sim5\%$) 
    are smaller than the symbols. The ratios at all pseudorapidities are calculated for the average charge $\frac{h^+ + h^-}{2}$.   }
\label{fig3}
\end{figure}

Because these ratios are constructed with events from the same data run, many corrections cancel out. 
The only correction 
that was applied to these ratios is related to trigger inefficiencies that become important in peripheral 
events. The 
dominant systematic error in these ratios stems from the determination of the average number of binary 
collisions in
each centrality data sample. This error is shown as a shaded band around 1 in Fig. \ref{fig3}. 

\begin{table}[hb] 
\vspace*{-12pt}
\caption[]{$N_{part}$ and $N_{coll}$ values extracted from HIJING calculations for d+Au collision}\label{tab2}
\vspace*{-14pt}
\begin{center}
\begin{tabular}{lccc}
\hline\\[-10pt]
$Centrality$ & $N_{part}(Au)$ & $N_{part}(d)$ & $N_{coll}$ \cr
 
\hline\\[-10pt]
$Central 0-20\%$ & 12.5  & 1.96  & 13.6 $\pm$ 0.3 \cr 

$Semi-central 30-50\%$ & 7.36 & 1.79 & 7.9 $\pm$ 0.4 \cr 

$Peripheral 60-80\%$ & 3.16 & 1.39 &$ 3.3 \pm 0.4$    \cr 

\hline 
\end{tabular}
\end{center}
\end{table}

The four panels of figure \ref{fig3} show the central $R^{central}_{CP}$ (filled symbols) and semi-central $R^{s
emi-central}_{CP}$
(open symbols) ratios for the four $\eta$ settings. At $\eta=0$ on the left-most panel, the 
central events yields are 
systematically higher than those of the semi-central events, and at the right-most panel ($\eta = 3.2) $ the 
trend is reversed; the yields of central events are 
$\sim 60\%$ lower than those from semi-central events at all values of transverse momenta.
 
\section{Discussion}\label{discussion}

Figures \ref{fig:ratio} and \ref{fig3} can be described in the following way: at $\eta=0$ the Cronin like enhancement is 
produced by coherent multiple scatterings of deuteron valence quarks on a saturated Au (for $ x \leq 0.01$).
The height of the enhancement increases with the centrality of the collisions because the number of 
scatterings is higher. As one moves to a higher rapidity y, the probability of gluons emission grows as $ P \sim \alpha_s  y$ and additional corrections have to be included. These corrections can be written as:    
$   \frac{d N}{d (ln \frac{1}{x})}= \alpha_{S}(2N - N^2) $  
within saturated and non-linear systems or 
 $  \frac{d N}{d(ln \frac{1}{x})}= \alpha_{S} 2N  $ in linear non-saturated systems. The variable N in these equations is related to the density 
of gluons in the nuclei. The linear term on the right side of these equations 
describes the emission of gluons and the quadratic term represents interactions 
between gluons that reduce their numbers.
The numerator of the nuclear modification factor $R_{dAu}$ is growing slowly with rapidity because of 
the taming effect of the quadratic term in the equation mentioned above, while the denominator 
continues to grow because the p+p system is more dilute. A similar suppression at all $p_{T}$ is seen
in Figure \ref{fig:ratio}. The strength of that suppression is proportional to the number of participant nucleons in the gold ion because it implies a higher number of gluons in the system making the effect of the quadratic term more and more important. This effect is also seen  in Figure \ref{fig3} as the pseudorapidity of the detected particles changes from 0 to 3.2 the suppression is stronger for the central sample of events. 

A recent calculation of the nuclear modification factor $R_{dAu}$ based on a description of 
the Au wave function as a color glass condensate and the deuteron as a dilute system of 
valence quarks \cite{latestJamal} agrees well with the BRAHMS results at $\eta = 3.2$.

\section*{Acknowledgment(s)}
This work was supported by 
the Office of Nuclear Physics of the U.S. Department of Energy, 
the Danish Natural Science Research Council, 
the Research Council of Norway, 
the Polish State Committee for Scientific Research (KBN) 
and the Romanian Ministry of Research.

%\section*{Note(s)} (footnotes will be typeset in this way) 
%\begin{notes}
%\item[a]
%Permanent address: Institute, City, Country;\\ 
%E-mail: author@host.domain.name
%\end{notes}

\vfill\eject
\end{document}